\pdfoutput=1
\documentclass[11pt,a4paper]{article}
\usepackage{amsmath,amssymb}
\usepackage{braket}
\usepackage{url}
\usepackage{mathtools}
\usepackage{amsmath}              
\usepackage{graphics,graphicx,epsfig,ulem} 
\usepackage{subfigure,subfig}
\usepackage{feynmp}
\usepackage{slashed}

\DeclareMathAlphabet{\mathbold}{U}{zeur}{b}{n}

\renewcommand\[{\left[}
\renewcommand\]{\right]}

\def\beq{\begin{equation}}
\def\eeq{\end{equation}}
\def\[{\begin{equation}}
\def\]{\end{equation}}

\newcommand{\startappendix}{
\setcounter{section}{0}
\renewcommand{\thesection}{\Alph{section}}
\renewcommand{\theequation}{\Alph{section}.\arabic{equation}}}

\newcommand{\Appendix}[1]{
\refstepcounter{section}
\begin{flushleft}
{\Large\bf Appendix: #1}
\end{flushleft}}

\begin{document}
\numberwithin{equation}{section}

\title{
\vspace{2.5cm}
\Large{\textbf{Perturbative growth of high-multiplicity W, Z and Higgs production
processes at high energies}}}

\author{Valentin V. Khoze\footnote{valya.khoze@durham.ac.uk}\\[4ex]
  \small{\it Institute for Particle Physics Phenomenology, Department of Physics} \\
  \small{\it Durham University, Durham DH1 3LE, United Kingdom}\\[0.8ex]
}

\date{}
\maketitle

\begin{abstract}
 \noindent Using the classical recursion relations we compute  scattering amplitudes 
 in a spontaneously broken Gauge--Higgs theory into final states involving high multiplicities of  
 massive vector bosons and Higgs bosons. These amplitudes are computed in the kinematic regime where the number
 of external particles $n$ is $ \gg 1$ and their momenta are non-relativistic. Our results generalise the previously
 known expressions for the amplitudes on the multi-particle thresholds to a more non-trivial kinematic domain.
 We find that the amplitudes in spontaneously broken gauge theories grow factorially with the numbers of particles produced, and that
this factorial growth is only mildly affected by the energy-dependent formfactor computed in the non-relativistic limit. 
This is reminiscent of the behaviour previously found in massive scalar theories. 
Cross sections are obtained by integrating the amplitudes squared over the non-relativistic
phase-space and found to grow exponentially at energy scales in a few hundred TeV range if we use the non-relativistic 
high multiplicity limit. This signals a breakdown of perturbation theory and indicates that the weak sector of the Standard
Model becomes effectively strongly coupled at these multiplicities. There are
interesting implications for the next generation of hadron colliders both for searches of new physics phenomena beyond
and within the Standard Model.
\end{abstract}

\bigskip
\thispagestyle{empty}
\setcounter{page}{0}

\newpage


\section{Introduction}\label{sec:intro}

The problem of divergences affecting large orders in perturbation theory is well known \cite{Dyson,Lipatov:1976ny,Brezin:1976vw,'tHooft:1977am},
and is often seen as an academic problem which reflects the asymptotic nature of perturbative series. 
This problem however is brought to an entirely new level when the perturbation theory breakdown is realised already at the leading order.
The physical quantities of interest in this case are associated with the scattering processes involving high multiplicities $n$ of particles produced in the final state
in the $n \gg 1$ limit.
At sufficiently high energies the production of such high multiplicity final states with $n$ greater than the inverse coupling constant, becomes kinematically allowed 
and the $n$-point scattering amplitudes can become large already at leading order -- i.e. tree level in a weakly coupled theory.

The motivation of this paper is to study the behaviour of scattering processes involving large numbers massive vector bosons and Higgs bosons
produced at high energy collisions. The underlying model is a spontaneously broken gauge theory, and the amplitudes will be computed 
on and off the multi-particle threshold, thus generalising the previously available results for the on-threshold amplitudes in \cite{VVK}.

In the case of the $\phi^4$-type scalar field theories, there is a direct link between the number of contributing Feynman diagrams, 
which grows as $n!$ at large $n$, and the and the overall expressions for the scattering amplitudes ${\cal A}_{n}$.
Multi-particle amplitudes in scalar theory were studied in detail in the 90s and were found to exhibit factorial growth
leading to the ultimate breakdown of the standard weakly-coupled perturbation theory, as reviewed in \cite{Voloshin:1994yp,LRT}, see also
 Refs.~\cite{Cornwall,Goldberg,Voloshin64,Argyres73,Brown,Voloshin80,LRST,Son}. However, this direct connection 
 between the on-shell quantities and the number of Feynman diagrams does not hold 
 in gauge theory because of the cancellations between individual diagrams which is a consequence of gauge invariance and the on-shell conditions. 
 Nevertheless, it was shown in  \cite{VVK} that in the case of spontaneously broken gauge theories, the amplitudes still grow
 factorially with the numbers of emitted Higgses as well as massive vector bosons. 
 
There are two objectives for our study: one is to demonstrate that the breakdown of perturbation theory previously found 
in scalar QFTs also applies to the weak sector of the Standard Model  -- or more precisely in any spontaneously broken gauge theory, and for
amplitudes in a generic non-relativistic kinematics.
Our second point is to emphasise that the energies and multiplicities involved are not unreachable with future particle experiments. 
There is an exciting possibility that these processes can be probed at the next generation of hadron colliders. 
The simple estimates carried out in this paper, however, will assume a conservative non-relativistic limit which is not well-suited
for drawing phenomenological conclusions about the lower limit on the energy scale where the weakly coupled perturbation theory would become
strong.

Another motivation for studying the high-multiplicity production in the electroweak sector at high energies is their analogy and a
potential complementarity with
the  topologically non-trivial transitions over the sphaleron barrier \cite{Ringwald,  KR, KRTetc, Mattis} which 
violate the Baryon plus Lepton ($B+L$) number in the SM. The common point between these two types of processes is the 
high multiplicities of the vector bosons and Higgs particles in the final state. In both cases this is the regime where ordinary perturbation
theory breaks down.

The paper is organised as follows. In the following section we will describe the method for computing the amplitudes on and off
the multi-particle threshold in the double-scaling high-multiplicity low-kinetic-energy limit we introduce. This technique is first applied to
a scalar theory with a non-trivial vacuum expectation value of the field. In section 3 this method is applied to our main case of interest -- the 
Gauge-Higgs theory. The phase-space integration is described in section 4, and our conclusions are presented in section 5. 
Our main results for the amplitudes are in Eqs.~\eqref{Ah-finkn}-\eqref{AZ-finkn}, and the cross section in the non-relativistic limit
is given in Eqs.~\eqref{eq:FnmH2},\eqref{eq:FnmH3}.

\medskip
\section{Summing tree graphs on and off the multi-particle threshold}
\label{sec:2}
\medskip

Our approach for computing tree-level high-multiplicity amplitudes in the Gauge-Higgs theory
is based on solving recursion relations between the $n$-point amplitudes involving massive vector and Higgs bosons
for different values of $n$. In the high-multiplicity regime, the outgoing particles will have non-relativistic velocities as they
are produced not far above their mass thresholds, thus we can simplify the amplitudes recursion relations by assuming
the non-relativistic limit. 

The main features of the method are best
explained following a simple example of a single real scalar field $h(x)$ with non-vanishing VEV $\langle h \rangle = v$ 
\[
{\cal L}(h) \,= \, \frac{1}{2}\, \partial^\mu h \, \partial_\mu h\, -\,  \frac{\lambda}{4} \left( h^2 - v^2\right)^2
\,.
\label{eq:LSSB}
\]
This scalar theory with the spontaneously broken $h\to -h$
  ${\cal Z}_2$ symmetry can be seen as a simplified version of the
Higgs sector of the SM in the unitary gauge.  
%
In the following
section we will apply this approach to our main case of interest -- the Gauge-Higgs theory. The goal of the present section 
is to derive a simple prescription for writing down the relevant recursion relations between the amplitudes in the example
of a relatively straightforward scalar QFT case \eqref{eq:LSSB}. Our results
for the scalar theory \eqref{eq:LSSB} are new, while our approach is similar to Ref.~\cite{LRST} 
where the unbroken $\phi^4$ theory was considered. 

Tree-level amplitudes for production of $n$ bosons from a virtual single-boson state, ${\cal A}_{n}:={\cal A}_{1\to n} $, 
are classical objects (no loops, $\hbar\to 0$), it is well-known that their generating functionals satisfies classical equations of motion 
with an external source. By differentiating $n$ times with respect to the source and setting it to zero, one obtains the recursion recursion
relations for the tree amplitudes, or currents, see e.g. \cite{Boulware:1968zz,Berends:1987me}.
We introduce the physical VEV-less scalar $\varphi(x) = h(x)-v$,
describing bosons of mass $M_h = \sqrt{2\lambda}\,v$. It satisfies the classical equation arising from \eqref{eq:LSSB},
\[
-\, (\partial^\mu \partial_\mu +M_h^2)\, \varphi \,=\, 3\lambda v\, \varphi^2 \,+\,  \lambda\,\varphi^3.
\label{eq:varphi}
\]
This classical equation in momentum space allows one to read-off directly the structure of the recursion relation for tree-level scattering amplitudes as follows:
\begin{eqnarray}
&& (P_{\rm in}^2 -M_h^2)\, {\cal A}_{n} (p_1\ldots p_n) \,=\, 3 \lambda v \sum_{n_1,n_2}^n  \delta^n_{n_1 +n_2} \sum_{\cal P}
 {\cal A}_{n_1} (p^{(1)}_1,\ldots,p^{(1)}_{n_1}) \, {\cal A}_{n_2} (p^{(2)}_1\ldots p^{(2)}_{n_2}) \nonumber \\
&& \qquad +  \lambda  \sum_{n_1,n_2, n_3}^n  \delta^n_{n_1 +n_2+n_3} \sum_{\cal P}
 {\cal A}_{n_1} (p^{(1)}_1\ldots p^{(1)}_{n_1}) \, {\cal A}_{n_2} (p^{(2)}_1\ldots p^{(2)}_{n_2}) \, {\cal A}_{n_3} (p^{(3)}_1\ldots p^{(3)}_{n_2})
\,.
\label{eq:Amps}
\end{eqnarray}
Here $P_{\rm in}^\mu = \sum_{i=1}^n p_i^\mu$ is the incoming momentum, and 
the sums over ${\cal P}$ involve permutations across the  sets of momenta $\{ p^{(1)}_i\}$,  $\{ p^{(2)}_i\}$, and/or
$\{ p^{(1)}_i\}$,  $\{ p^{(2)}_i\}$ and $\{ p^{(3)}_i\}$ of the the individual amplitudes on the right hand side of Eq.~\eqref{eq:Amps}.

The special case of \eqref{eq:Amps} where all outgoing $n$ particles are produced on their mass threshold, 
i.e. with vanishing spatial momenta $\vec{p}_i \equiv 0$, is particularly simple. 
Here the amplitudes are constants, the kinematics is trivial and one can sum over the permutations with the result,
\[
M_h^2(n^2 -1)\, {\cal A}_{n}  \,=\, 3 \lambda v \sum_{n_1,n_2}^n  \delta^n_{n_1 +n_2} \frac{n!}{n_1! n_2!}
 {\cal A}_{n_1}  \, {\cal A}_{n_2}  +
\lambda \sum_{n_1,n_2, n_3}^n  \delta^n_{n_1 +n_2+n_3} \frac{n!}{n_1! n_2! n_3!}
 {\cal A}_{n_1} \, {\cal A}_{n_2}  \, {\cal A}_{n_3} 
\,.
\label{eq:Amps2}
\]
The solution of  this momentum-independent recursion relation can be captured and recast in terms of the
the amplitudes generating function which solves the original Euler-Lagrange equation  \eqref{eq:varphi}
for the $\vec{x}$-independent field $\varphi(t)$
with the initial condition \cite{Brown},
\[
\varphi(t)\,=\, z(t) + {\cal O}(z^2)\,,
\quad {\rm where} \quad
z(t)\,:=\,z_0 \, e^{iM_h t}\, = \, z_0 \, e^{i\sqrt{2\lambda}\,v\,t}\,.
\label{eq:leadz}
\]
The generating function for the on-shell amplitudes 
is a complex-valued solution of  \eqref{eq:varphi} which contains only the positive-energy harmonics, $e^{inM_h t}$.
For a fixed $n$ each plain wave describes the $n$-particle final state
at rest with $n$ bosons of mass $M_h=\sqrt{2\lambda}\,v$. 
Hence the generating function is a holomorphic function of the complex argument $z$,
\[
\varphi (z) = \sum_{n=1}^\infty d_n\, z^n\,, \quad {\rm with} \quad d_1=1\,,
\label{eq:holos}
\]
with the individual $n$-point amplitudes on the multi-particle threshold given by,
\[
{\cal A}_{ n}\,=\, 
\left.\left(\frac{\partial}{\partial z}\right)^n \varphi (z)\,\right|_{z=0}
\,=\, n!\, d_n
\,.
\label{eq:amplnhd}
\]
Substituting the Taylor expansion \eqref{eq:holos} into the classical equation \eqref{eq:varphi}, one immediately
finds the recursion relation between the coefficients $d_n$,
\[
(n^2 -1)\, d_{n}  \,=\, 3 \, \frac{\lambda v}{M_h^2} \sum_{n_1,n_2}^n  \delta^n_{n_1 +n_2} 
 d_{n_1}  \, d_{n_2}  +
\frac{\lambda}{M_h^2} \sum_{n_1,n_2, n_3}^n  \delta^n_{n_1 +n_2+n_3} 
 d_{n_1} \, d_{n_2}  \, d_{n_3} \,,
\label{eq:Amps3N}
\]
which, as expected, is equivalent to Eq.~\eqref{eq:Amps2} (note that there are no $n!$ factors when the recursion relations are expressed 
in terms of the Taylor coefficients $d_n$, rather than ${\cal A}_{ n}$).
The  generating function $\varphi(z)$ which solves \eqref{eq:varphi}, is known analytically\footnote{In fact $h(z) = v +\varphi(z)$  
in \eqref{sol-SSB} corresponds  the well-known kink solution in Euclidean time which interpolates between the two vacua at $h=\pm v$.}
 \cite{Brown},
\[
\varphi  \,=\, \frac{z}{1-{z}/({2v})} \, , \quad {\rm hence} \quad
d_n\,=\,  (2v)^{1-n}   \, , \quad {\rm and} \quad
{\cal A}_{ n}\,=\,n!\, (2v)^{1-n}
\label{sol-SSB}
\]
The main lesson of this exercise was to show that our tree-level threshold amplitudes ${\cal A}_{1\to n}^{\rm threshold}$  grow factorially with the number of final particles $n$. However, our main concern are the amplitudes above the threshold, and so we need to recover their
dependence on the kinematics of the final state.


Away from the multi-particle threshold, the external particles 3-momenta $\vec{p}_i$ are non-vanishing and in the non-relativistic limit which we will adopt,
they are small compared to the particle masses.
In this limit we can characterise the process in the COM frame by the non-relativistic kinetic energy $E_n^{\rm \, kin}$ of the final particles.
In general kinematics, the amplitudes are determined by the recursion relation \eqref{eq:Amps} which we need to solve.
Amplitudes {\it on} the threshold were already found in \eqref{sol-SSB}; they correspond to $E_n^{\rm \, kin}=0$. At small values of 
$E_n^{\rm \, kin}/(n M_h)$
the leading correction to these amplitudes turns out to be proportional to $E_n^{\rm \, kin}$ itself. This fact  
is simply the consequence of the permutation symmetry 
acting on the particle momenta $\vec{p}_i$ and of the Galilean invariance of the amplitude \cite{LRST}. 
Hence, the scattering amplitude in the
non-relativistic approximation takes the form:
\[
{\cal A}_{n} (p_1 \ldots p_n) \,=\, {\cal A}_{ n} + \, {\cal M}_n \, E_n^{\rm \, kin} \,:=\, {\cal A}_{ n}
+ \, {\cal M}_n \, n\, \varepsilon\,,
\label{eg:epsexp}
\]
where ${\cal A}_{ n}$ is the threshold amplitude, ${\cal M}_n \, n\, \varepsilon$ denotes the leading order momentum-dependent
contribution, and
$\epsilon$ is the kinetic energy per particle per mass,
\[
\varepsilon = \frac{1}{n\, M_h}\, E_n^{\rm \, kin}\, =\, 
\frac{1}{n}\,  \frac{1}{2 M_h^2} \, \sum_{i=1}^n \vec{p}_i^{\, \, 2} \,.
\]
In the non-relativistic limit we have $\varepsilon \ll 1$. 


Working in the CoM frame the incoming momentum is $P_{\rm in}^\mu= ( n M_h (1+\epsilon), \vec{0})$ and the left hand side
of the recursion relation \eqref{eq:Amps} takes the form, 
$M_h^2 \left( n^2(1+\varepsilon)^2-1 \right) \, {\cal A}_{n} (p_1\ldots p_n)$.
Using the relation \eqref{eg:epsexp} and working
at the order-$\varepsilon^1$, the recursion relations \eqref{eq:Amps} amount to:
\begin{eqnarray}
\left( n^2-1 \right) n\, \varepsilon\, {\cal M}_n \,+\, 2 n^2\, \varepsilon\,{\cal A}_{ n}
&=&  6\, \frac{\lambda v}{M_h^2} \,\sum_{n_1,n_2}^n  \delta^n_{n_1 +n_2} \sum_{\cal P}
 E_{n_1}^{\rm \, kin} \,{\cal M}_{n_1} \, {\cal A}_{n_2}  
 \label{eq:AmpsNew}
  \\
&&+\, 3 \,\frac{\lambda} {M_h^2} \sum_{n_1,n_2, n_3}^n  \delta^n_{n_1 +n_2+n_3} \sum_{\cal P}
E_{n_1}^{\rm \, kin} \,{\cal M}_{n_1} \, {\cal A}_{n_2} \, {\cal A}_{n_3} \,.
\nonumber
\end{eqnarray}
This is the equation for ${\cal M}_n $.
On its right hand side we have used the notation $E_{n_1}^{\rm \, kin}$ (rather than e.g. $n_1 \varepsilon$) to denote the 
total kinetic energy of the outgoing particles of the sub-process $1 \to n_1$. This quantity is defined by
$E_{n_1}^{\rm \, kin} := \frac{1}{2 M_h} \sum_{i=1}^{n_1} \left(\vec{p}_i - \frac{1}{n_1} \vec{p}_0\right)^2,$ 
where we have taking into account
 that the initial state of this sub-process
is no longer at rest, $\vec{p}_0 := \sum_{i=1}^{n_1} \vec{p}_i \, \neq \vec{0}$. 
Hence,
\[
 E_{n_1}^{\rm \, kin}  =\, \frac{1}{2 M_h}\left[ \sum_{i=1}^{n_1} \vec{p}_i^{\,2} - \frac{1}{n_1} \vec{p}_0^{\, 2}\right]=
\frac{1}{2 M_h}\left[\left(1- \frac{1}{n_1} \right) \sum_{i=1}^{n_1} \vec{p}_i^{\,2} - \frac{1}{n_1}\sum_{i=1}^{n_1}\sum_{j\neq i}^{n_1}  \vec{p}_i  \vec{p}_j \right]
\]
To simplify this, we note that all $n$ 
 external momenta 
$\vec{p}_1 \ldots \vec{p_n}$ will contribute in the sums on the right hand side of the recursion relation \eqref{eq:AmpsNew}. This allows us to
effectively include all $n$ momenta in the double sum above and use
the substitution, 
\[
\sum_{i=1}^{n_1}\sum_{j\neq i}^{n_1}  \vec{p}_i  \vec{p}_j \Longrightarrow 
\frac{n_1(n_1-1)}{n(n-1)} \sum_{i=1}^{n}\sum_{j\neq i}^{n}  \vec{p}_i  \vec{p}_j =\,
-\,\frac{n_1(n_1-1)}{n(n-1)} \sum_{i=1}^{n}  \vec{p}_i^{\,2} \,.
\]
This gives the expression for the kinetic energy,
\[
E_{n_1}^{\rm \, kin}\,  \Longrightarrow\,  \frac{n\,(n_1-1)}{n-1}\,\, \varepsilon\,,
\label{KinEsubs}
\]
which we can use on right hand side of \eqref{eq:AmpsNew}.

We will also use the amplitude's coefficients $d_n$ and $f_n$ defined via:
\[
{\cal A}_{n} (p_1 \ldots p_n)  \,=\, n!\, (d_n \,+\, f_n \, \varepsilon)
\,,
\label{eq:2-16}
\]
rather than the amplitudes ${\cal A}_n$ and ${\cal M}_n $ in \eqref{eg:epsexp}).
The expression \eqref{eq:Amps3N} is the recursion relation  at the order-$\varepsilon^0$, 
and at the order-$\varepsilon^1$ from Eqs.~\eqref{eq:AmpsNew} and \eqref{KinEsubs} we have:
\begin{eqnarray}
\frac{n-1}{n}\left(
\left( n^2-1 \right)  f_n \,+\, 2 n^2\,d_{ n} \right)
&=&  6\, \frac{\lambda v}{M_h^2} \,\sum_{n_1,n_2}^n  \delta^n_{n_1 +n_2} 
\frac{n_1-1}{n_1}\,
 f_{n_1} \, d_{n_2}  
 \label{eq:AmpsNeps}
  \\
&&+\, 3 \,\frac{\lambda} {M_h^2} \sum_{n_1,n_2, n_3}^n \delta^n_{n_1 +n_2+n_3} 
\frac{n_1-1}{n_1}\,
 f_{n_1}  \, d_{n_2} \, d_{n_3} \,.
\nonumber
\end{eqnarray}
The factors of $\frac{n-1}{n}$ and $\frac{n_1-1}{n_1}$ on the left and right hand sides of this equation arise from
the kinetic energy formula \eqref{KinEsubs}.

The recursion relation  \eqref{eq:Amps3N} for the amplitudes on the multi-particle threshold is solved as before 
({\it cf.} Eq.~\eqref{sol-SSB})
and determines the full set of the
$d_n$ coefficients, $d_n\,=\,  (2v)^{1-n}$.
We can now go ahead and solve the second recursion relation \eqref{eq:AmpsNeps} to determine 
the coefficients $f_n$. Our result  is 
\[
f_n\,=\, - \, \left(\frac{7}{6} \,n \,+\, \frac{1}{6} \, \frac{n}{n-1}\right)\,  d_n\, \,, \quad {\rm for \,\,all} \,\,\, n\ge 2\,.
\label{eq:resultf}
\]
This result is obtained by solving an ordinary differential equation (as will be explained below) by iterations with {\it Mathematica}.
The resulting amplitude  to the order-$\varepsilon$  is then given by 
\[
{\cal A}_n (p_1 \ldots p_n)  \,=\, n!\,  (2v)^{1-n}\,\left( 1-\frac{7}{6}\,n\, \varepsilon \,-\, \frac{1}{6} \, \frac{n}{n-1} \, \varepsilon \,+\, \ldots \right).
\]
There are corrections to this expression at higher orders in $\varepsilon$, but it holds to the order $\varepsilon^1$ for any value of 
$n$.

\medskip

An important observation, first made in \cite{LRST}, is that by exponentiating the order-$n\varepsilon$ contribution,
one obtains the expression for the amplitude
which solves the original recursion relation
\eqref{eq:Amps}
to all orders in $(n\varepsilon)^m$ in the large-$n$ non-relativistic limit,
\[
{\cal A}_n (p_1 \ldots p_n)  \,=\, n!\,  (2v)^{1-n}\,\exp\left[-\frac{7}{6}\,n\, \varepsilon \right]\,,
\quad n\to \infty\,, \quad \varepsilon \to 0\,, \quad n\varepsilon = {\rm fixed}\,.
\label{eq:expsc}
\]
Simple corrections of order $\varepsilon$, with coefficients that are not-enhanced by $n$ are expected,
but the expression on right hand side of \eqref{eq:expsc} is correct to all orders $n\varepsilon$ in the double scaling
large-$n$ limit.
This observation follows from the fact that the exponential factor in \eqref{eq:expsc} can be absorbed into the $z$ variable
so that the expression in \eqref{eq:holos} with the rescaled $z$ on the right hand side,
\[
\varphi (z) = \sum_{n=1}^\infty d_n\, \left(z\, e^{-\frac{7}{6}\, \varepsilon}\right)^n\,, 
\label{eq:holos2}
\]
remains a solution to the classical equation
\[
-\, (d_t^2 +M_h^2)\, \varphi \,=\, 3\lambda v\, \varphi^2 \,+\,  \lambda\,\varphi^3.
\label{eq:varphi22}
\]
This implies that the individual $n$-point amplitudes will satisfy the recursion relations 
\eqref{eq:Amps2} and by the same token the general-kinematics recursion \eqref{eq:Amps}.
This is because the operator $P_{\rm in}^2 -M_h^2$ on the left hand side of (2.3)
becomes $M_h^2 \left( n^2(1+\varepsilon)^2-1 \right) =  M_h^2 \left( n^2-1 \right)$ in the
$n\to \infty$, $\varepsilon \to 0$ double-scaling limit.
Then as soon as \eqref{eq:holos2} satisfies \eqref{eq:Amps2}, it also satisfies the general-kinematics recursions \eqref{eq:Amps}
in the double-scaling limit.

Note that the exponentiated expression \eqref{eq:expsc} solves \eqref{eq:varphi22} for any constant factor in the exponent, but 
having solved the order-$\varepsilon$ recursions explicitly we have determined 
in \eqref{eq:resultf} the value of the constant to be $=-7/6$.\footnote{In the 
$\phi^4$ theory with no spontaneous symmetry breaking  
the authors of Ref.~\cite{LRST} have shown that the amplitudes scale with energy as
$\propto \,n!\, \exp\left[-\frac{5}{6}\,n\, \varepsilon \right].$ We will re-derive their result in the Appendix.}

\medskip

{\it Rescaled variables:} 
 To simplify the form of the recursion relations  -- especially in view of the applications to the coupled Gauge-Higgs
 equations in the following section -- 
 we define new rescaled dimensionless variables: 
 \[
t_{\rm new} =\, M_h\, t
\,, \quad 
z_{\rm new} =\, \frac{z}{2v}
\, \qquad {\rm and} \qquad  \phi := \, \frac{1}{2v} \, \varphi  \,=\, \sum_{n=1}^\infty {d}^{\rm new}_n \,z_{\rm new}^n
\,,
\label{eq:resc}
 \]
 so that ${d}^{\rm new}_n \, =\,  (2v)^{n-1} d_n$ which amounts to a particularly simple form of the recursive solution
 ${d}^{\rm new}_n \, =\, 1$ or all $n=1,2, \ldots \infty$ in Eq.~\eqref{sol-SSB},
 and is the reason why we introduced factors of 2 in the rescaling \eqref{eq:resc}.

 In terms of these new variables (and suppressing the superscript `new') the classical equation \eqref{eq:varphi22} takes the form
 \[
-(d_t^2 +1) \phi \,=\, 3 \phi^2 + 2\phi^3
\,,
\label{cleqphi}
\]
and  the amplitudes on the multi-particle threshold ({\it cf.} \eqref{eq:amplnhd}) are given by:
\[
{\cal A}_{ n}\,=\, (2v)^{1-n}\,
\left.\left(\frac{\partial}{\partial z}\right)^n \phi (z)\,\right|_{z=0}
\,=\, n!\, (2v)^{1-n}\, d_n
\,,
\label{eq:amplnhd2}
\]
where
\[\phi(z)  \,=\, \sum_{n=1}^\infty {d}_n \,z^n \,,\,\,\,{\rm and} \,\,\,\,
d_n \equiv 1 \,, \,\,\, n=1,2,\ldots,\infty
\,.
\label{eq:scfin}
\]

With the equation \eqref{cleqphi} defining the generating function for amplitudes on the multi-particle threshold,
the order-$\varepsilon$ correction to the generating function (in our rescaled variables) is determined by the differential equation
({\it cf.} \eqref{eq:AmpsNeps}):
\[
\frac{n-1}{n}\left(
\left( n^2-1 \right)  f_n \,+\, 2 n^2\,d_{ n} \right)
\,=\, 6\, \left( F \phi \,+\, F \phi^2\right) |_{z^n}\,,
 \label{eq:AmpsNepsR}
\]
where we defined the new function
\[
F(z) \,=\, \sum_{n=2}^\infty\, \frac{n-1}{n} \, f_n \, z^n\,.
\label{eq:Fz}
\]
Solving Eq.~\eqref{eq:AmpsNepsR} by iterations with {\it Mathematica} gives the $f_n$ coefficients in \eqref{eq:resultf}, which amounts to the
exponentiated form for the amplitude off the multiparticle mass-shell in the non-relativistic limit $\varepsilon \to 0$, with $n \varepsilon=$fixed,
\[
{\cal A}_n (p_1 \ldots p_n)  \,=\, n!\,  (2v)^{1-n}\,\exp\left[-\frac{7}{6}\,n\, \varepsilon \right]
\,.
\]
This is our main result for the tree-level high-multiplicity amplitudes in the scalar theory with SSB in the  $\varepsilon \to 0$, with $n \varepsilon=$ fixed
limit.

\medskip
\section{Multiparticle production in the Gauge-Higgs theory}
\label{sec:3}
\medskip

We are now ready to consider our main case of interest -- the electroweak sector of the Standard Model.
In the limit of the vanishing mixing angle $\theta_{\rm W}$, the weak interactions are described by
the SU(2) gauge theory spontaneously broken by the vacuum expectation value $v$ of the Higgs doublet,
\[
{\cal L}\,=\, -\frac{1}{4} F^{a\,\mu\nu}F^a_{\mu\nu} \,+\, |D_\mu H|^2 \,-\, 
\lambda\left(|H|^2-\frac{v^2}{2}\right)^2\,.
\label{eq:LGH}
\]
We adopt the standard unitary gauge where the Goldstone bosons are gauged away, and the Higgs doublet is described 
by a single real scalar $h(x)$,
\[
H= \frac{1}{\sqrt{2}} \left(0,h\right),\,
\]
The Higgs potential in terms of $h$ takes the same form as in Eq.~\eqref{eq:LSSB}. 
The particle content of the model is given by the neutral Higgs state,
$h$, and a triplet of massive vector bosons, $W^{\pm}$ and $Z^0$,
described by $A_\mu^a$ with $a=1,2,3$, which we will collectively refer to as $V$. The Higgs mass and the mass of the vector boson triplet are given by,
\[
M_h\,=\, \sqrt{2\lambda}\,v \,\simeq\, 125.66\, {\rm GeV}\,, \qquad
M_V\,=\, \frac{g v}{2}\,\simeq\, 80.384 \, {\rm GeV}\,,
\label{eq:masses}
\]
where we have also shown their numerical values, set by the SM Higgs and $W$ boson masses, which will be uses in our calculations of the amplitudes below.

We want to study the processes where colliding protons first produce an
intermediate virtual state, which can be either the Higgs or a gauge boson. This intermediate highly virtual boson then decays into 
$n$ Higgs bosons and $m$ vector bosons $1 \to n+m$, which is the multi-particle production process we concentrate upon.
The multiplicity of the final state $n+m$ is assumed to be large so that most of the
energy carried by the virtual state is spent to achieve the multi-particle mass threshold for the $n+m$ final particles. Above the threshold,
the momenta of the final state particles are assumed to be non-relativistic. It is convenient to describe the kinematics working in
the Lorentz frame where the initial virtual boson is at rest. In this frame,
\[
P_{\rm in}^\mu=\,(P_{\rm in}^0, \vec{0}) \,= \,\sum_{j=1}^{n} p_j^\mu\,+\,\sum_{k=1}^{m} p_{k}^\mu
\,,
\label{eqn:thrp1}
\]
where the first sum on the right hand side is over the $n$ Higgs bosons, and the second sum is over the $m$ 
vector bosons produced in the final state.
We will make an additional simplifying assumption that the total momentum is conserved separately in the Higgs,
and in the vector boson sectors, i.e.
\[
\sum_{j=1}^{n} \vec{p}_j \,=\, 0\,, \quad {\rm and}\quad \sum_{k=1}^{m} \vec{p}_k \,=\, 0\,.
\label{eqn:thrp2}
\]
In the rest frame of  $P_{\rm in}^\mu=\,(P_{\rm in}^0, \vec{0})$ this amounts to a single constraint on the overall kinematics 
of the multi-particle final state and imposing it should not affect the result of integrating over the $(n+m)$-particle
phase space in the $n+m \gg 1$ high-multiplicity limit.
With these considerations in mind we can thus express the initial virtual-state momentum in the form,
\[
P_{\rm in}^\mu=\,\left(nM_h(1+\varepsilon_h) + mM_V(1+\varepsilon_V)\, , \, \vec{0}\right) \,,
\label{eqn:thrp}
\]
where $\varepsilon_h$ and $\varepsilon_V$ denote the average non-relativistic kinetic energies of the
Higgs bosons, and of the vector bosons, per particle per mass,
\[
\varepsilon_h \,=\,\frac{1}{n}\,  \frac{1}{2 M_h^2} \, \sum_{j=1}^n \vec{p}_j^{\, \, 2} \,,
 \qquad
\varepsilon_V \,=\,\frac{1}{m}\,  \frac{1}{2 M_V^2} \, \sum_{k=1}^m \vec{p}_k^{\, \, 2} \,.
\label{eq:kinHV}
\]
In the non-relativistic limit for the final state we have $0\le \varepsilon_h \ll 1$ and $0\le \varepsilon_V \ll 1$.

\subsection{Recursion relations for amplitudes on the multi-particle threshold}

Following the approach of \cite{VVK} we 
will consider the amplitudes for processes with final states which do not contain transverse polarisations of the vector bosons,
and concentrate on the production of longitudinal polarisations, $A_L^a$ and Higgses $h$. The classical equations 
for spacialy-independent fields readily follow from the Lagrangian \eqref{eq:LGH} in the unitary gauge
(these are Eqs. (3.8)-(3.9) of \cite{VVK}),
\begin{eqnarray}
&&- \, d_t^2 h \,\, =\,   \lambda\,h^3  -\lambda v^2\,h +\frac{g^2}{4} (A_L^a)^2 h
\,,\label{cleq-h3}\\
&&-\,  d_t^2 A_L^a \,=\, \frac{g^2}{4} h^2 A_L^a \,.
\label{cleq-A3}
\end{eqnarray}
The generating function of the amplitudes on the multi-$h$, multi-$V_L$ threshold,
is the classical solution of this system of equations given by analytic functions of two variables,
\[
z(t) \,=\, z_0\, e^{i M_h t}\,,\quad {\rm and}\quad
w^a(t)\,=\, w_0^a\, e^{i M_V t}\,,
\label{eq:zw}
\]
with the leading-order terms being,
\[
h(t)\,=\, v + z(t) + \ldots \,, \quad {\rm and } \quad
A_{L}^a (t)\,=\, w^a(t) + \ldots\,.
\label{eqs:incHA}
\]
The double Taylor expansion for the two generating functions in terms of the $z$ and $w^a$ variables takes the form:
\begin{eqnarray}
h(z,w^a) &=& v\,+\, 2v\, \sum_{n=0}^{\infty}\sum_{k=0}^{\infty} \, d(n,2k)\,\left(\frac{z}{2v}\right)^n\,\left(\frac{w^a w^a}{(2v)^2}\right)^k
\,,\label{eq:dTh-fin}
\\
A_{L}^a(z,w^a) &=& w^a\,\sum_{n=0}^{\infty}\sum_{k=0}^{\infty} \, a(n,2k)\, \left(\frac{z}{2v}\right)^n\,\left(\frac{w^a w^a}{(2v)^2}\right)^k
\,,
\label{eq:dTA-fin}
\end{eqnarray}
where $z(t)$ and $w^a(t)$ are given by Eqs.~\eqref{eq:zw}, and the lowest-order Taylor coefficients are $d(0,0)=0$ and $a(0,0)=1$
in agreement with \eqref{eqs:incHA}. 
The explicit scaling factors of $2v$ are introduced on the right hand side of the above equations to maximally simplify the form of
the solutions for the Taylor expansion coefficients. In particular, in this notation 
we will have $d(n,0) = 1$ for all values of $n\ge 1$, in agreement with the solution of the scalar equation \eqref{eq:scfin}
in the previous section.

To simplify the form of the classical equations and to emphasise that they depend 
only on a single numerical parameter $\kappa$,
\[
\kappa :=\, \frac{g}{2\sqrt{2\lambda}} \,=\,  \frac{M_V}{M_h}\,, 
\label{eq:kappadef}
 \]
we introduce the rescaled dimensionless variables as in \eqref{eq:resc},
 \[
t_{\rm new} =\, M_h t
\,, \quad 
z_{\rm new} =\, \frac{z}{2v} =\, \frac{z_0}{2v}\, e^{i t_{\rm new}}\,,\quad 
w^a_{\rm new} =\, \frac{w^a}{2v} =\, \frac{w_0^a}{2v}\, e^{i \kappa t_{\rm new}}
\,, 
\label{eq:resctzw}
 \]
 and also define the dimensionless fields, $\phi$ for the VEVless scalar, and ${\rm A}$ for the vector bosons, via:
 \[
h=\, v\,(1+2\phi) \,, \quad 
A_L^a =\, w^a {\rm A} =\, 2v \,w^a_{\rm new} \, {\rm A}
\,.
\label{eq:rescHA}
 \]
Note that the vector boson configuration ${\rm A}$ on the right hand side of the second equation in 
\eqref{eq:rescHA} no longer contains the isospin index $a = 1,2,3$
which has been factored out into the $w^a$ prefactor. 
We use these new dimensionless variables and fields to re-write Eqs.~\eqref{cleq-h3}-\eqref{cleq-A3} in the form,
\begin{eqnarray}
-\, (d_t^2 +1) \phi &=& 3 \phi^2 + 2\phi^3 + 2 \kappa^2\left(1 +2\phi\right)(w^a w^a)\,{\rm A}^2
\,,\label{cleq-h3r}\\
-\, (d_t^2 +\kappa^2) \, w^a {\rm A} &=& 4\kappa^2 (\phi+\phi^2)\, w^a {\rm A} \,.
\label{cleq-A3r}
\end{eqnarray}
As expected, this system of equations depends on a single dimensionless parameter $\kappa$ and we note that 
in the $\kappa\to 0$ limit the
equation \eqref{cleq-h3r} reproduces the scalar-field equation \eqref{cleqphi} of section 2.
The Taylor expansions \eqref{eq:dTh-fin}-\eqref{eq:dTA-fin} are also simplified in terms of the rescaled variables,
\[
\phi \,=\,  \sum_{n=0}^{\infty}\sum_{k=0}^{\infty} \, d(n,2k)\, z^n\,W^k
\,, \quad
{\rm A} \,=\, \sum_{n=0}^{\infty}\sum_{k=0}^{\infty} \, a(n,2k)\, z^n\,W^k\,,
\label{eq:dThA-finF}
\]
where  we have introduced the squared $w$ variable,
\[ W=w^aw^a\,.
\]

The recursion relations for the coefficients $d(n,2k)$ and $a(n,2k)$ are obtained by substituting the Taylor expansions
\eqref{eq:dThA-finF} into the classical equations \eqref{cleq-h3r}-\eqref{cleq-A3r} and selecting the 
$z^n\, W^k$ monomials,
\begin{eqnarray}
\left[(n+2k\, \kappa )^2 -1\right]\, d(n,2k)  &=& \left.\left[
3 \phi^2 + 2\phi^3 +2\kappa^2\left(1 +2\phi\right)W {\rm A}^2
\right]\right|_{z^n W^k}\,,
\label{cleq-h3rT}\\
\left[(n+ \kappa+ 2k\, \kappa )^2 -\kappa^2\right]\, a(n,2k)  &=& \left.\left[
4\kappa^2 (\phi+\phi^2)\, {\rm A}
 \right]\right|_{z^n W^k}
 \,.
\label{cleq-A3rT}
\end{eqnarray}
These equations were solved in \cite{VVK} by iterations using the numerical value
of $\kappa= M_W/M_h= 80.384/125.66\simeq 0.6397$. 
First we set $k=0$ and solve the Higgs equations \eqref{cleq-h3rT}
for all values of $n\ge 1$ thus determining all coefficients\footnote{The solution is
$d(n,0)=1$ for all $n\ge 1$ which is in agreement with the pure scalar theory result in \eqref{eq:scfin}.}
 $d(n,0)$. Then we solve the $A$-equations \eqref{cleq-A3rT}
 for the coefficients $a(n,0)$ for each $n$.
 Next we set $k=1$, and solve equations \eqref{cleq-h3rT} for all $n$ to determine $d(n,2)$. Following this, the coefficients
 $a(n,2)$ are found by solving  \eqref{cleq-A3rT} at $k=1$ for all values of $n$. This procedure is repeated for all values of $k$.
  
\begin{figure}[t]
\begin{center}
\begin{tabular}{cc}
\hspace{-.4cm}
\includegraphics[width=0.5\textwidth]{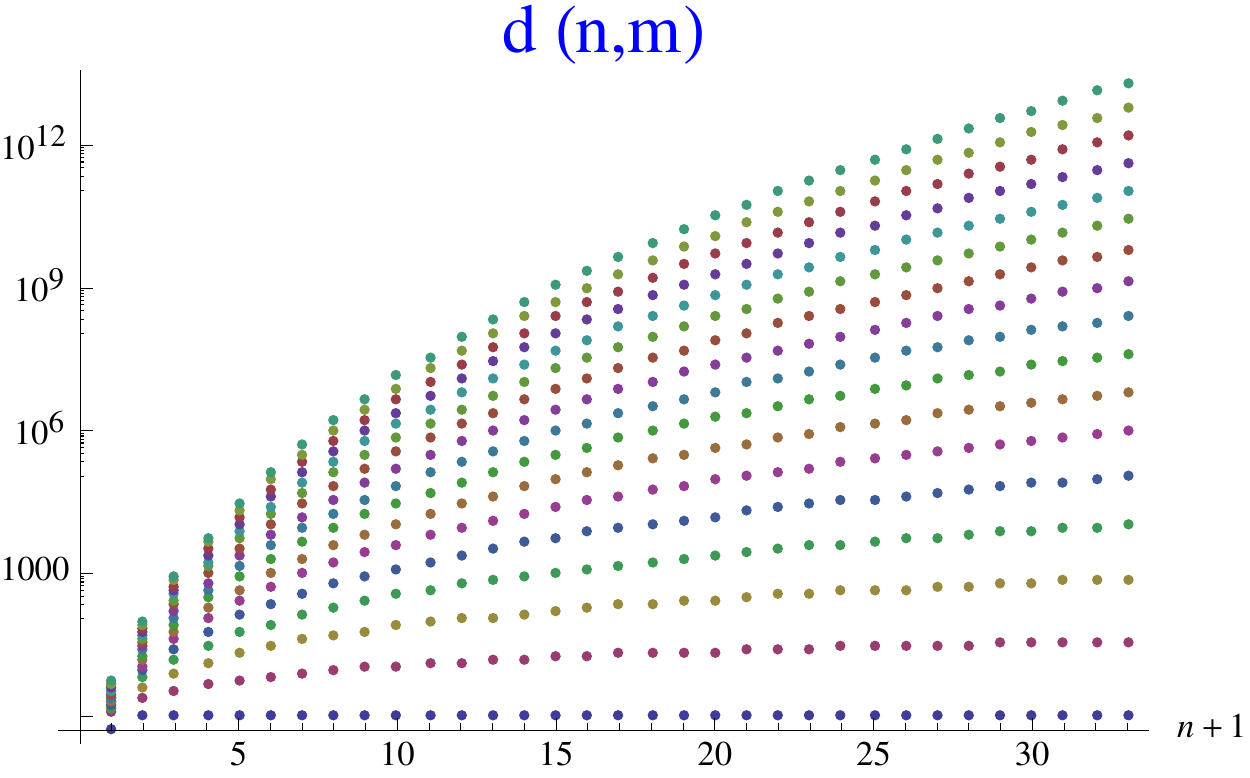}
&
\includegraphics[width=0.5\textwidth]{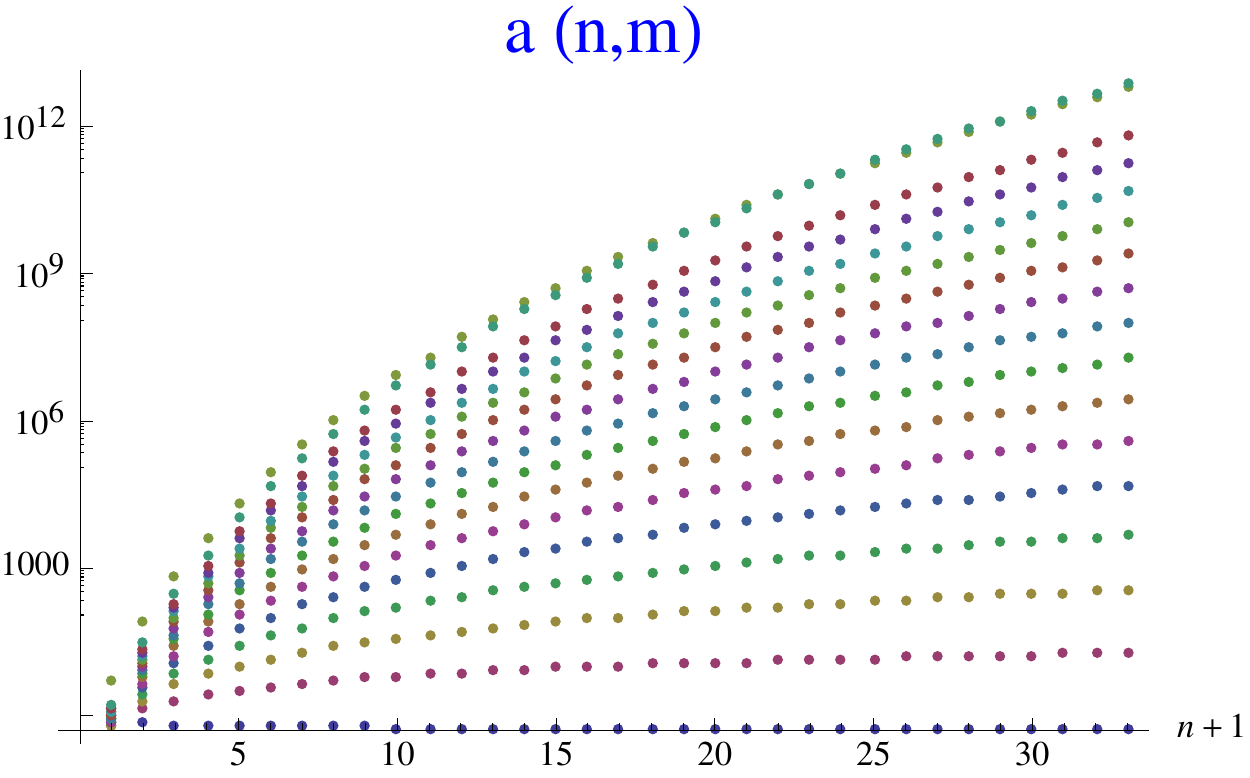}
\\
\end{tabular}
\end{center}
\vskip-.4cm
\caption{
Coefficients $d(n,m)$ and $a(n,m)$ for generating functions of amplitudes \eqref{Ah-fin}-\eqref{AZ-fin} at threshold, from
Ref.~\cite{VVK}.
The label $n=0,1,\ldots, 32$ is shown along the horizontal axis and the sequences of curves correspond to 
$m=0, 2, \ldots,32$ from bottom to top. $\kappa= M_W/M_h \simeq 0.6397$.
}
\label{fig:da}
\end{figure}

 This iterative algorithm was implemented in \cite{VVK} in {\it Mathematica}. One can solve for $d(n,2k)$ and $a(n,2k)$ to any desired values of
 $n$ and $k$ numerically.
  Tables 1-4 in Ref.~\cite{VVK} list numerical values of the coefficients\footnote{Note that in the notation of \cite{VVK} the value
  of $d(0,0)$ was $1/2$ while in the notation of the present paper $d(0,0)\equiv 0$. This is the consequence of working with the 
  scalar field $\phi$ shifted by the VEV rather than with $h$. The rest of the coefficients in the Tables 1-4 of \cite{VVK} are unchanged.}
  up to $d(32,32)$ and $a(32,32)$.
 In Figure~\ref{fig:da} we show the logarithmic 
 plots of all $d(n,m)$ and $a(n,m)$ for $n=0\ldots 32$ and $m=0, 2, \ldots,32$. These plots can be interpreted as sequences of curves, 
 each curve
 representing $d(n,m)$ and $a(n,m)$ as functions of $n$ for a fixed value of $m$. Increasing values of $m=0, 2, \ldots,32$ 
 corresponds to moving 
 upwards from lower to higher curves. 
   
 \medskip

The amplitudes on the multi-particle threshold are given by the following expressions
in terms of the Taylor expansion coefficients $d(n,2k)$ and $a(n,2k)$,
\[
  {\cal A}_{h^* \to n\times h + m\times Z_L}^{\rm (thr)}= \, (2v)^{1-n-m}\, n!\, m!\, d(n,m) \,,
  \label{Ah-fin}
 \] 
 and for the longitudinal Z decaying into $n$ Higgses and $m+1$ vector bosons we have,
 \[
 {\cal A}_{Z_L^* \to n\times h + (m+1)\times Z_L}^{\rm (thr)}= \, \frac{1}{(2v)^{n+m}}\, n!\, (m+1)!\, a(n,m)\,.
  \label{AZ-fin}
 \]
 The amplitudes with all varieties of $W^{\pm}_L$ and $Z_L$ in the final state, one should simply differentiate 
 with respect to $w^a$ with the appropriate values of the isospin index $a=1,2,3$.
 
 At $m=0$ the coefficients $d(n,0)=1$ for all $n\ge 1$ provide a useful reference point.
 After switching on $m >0$, the coefficients of the generating functions grow steadily with $m$, reaching $d(n,m) \sim 10^8$
 at $m\ge 16$ and $n \ge 27$; and $d(n,m) \sim 10^{13}$
 at $m=32 $ and $n =31$ 
 and similar growth with $m$ occurs for the $a(n,m)$ coefficients of the gauge field generating function
 \cite{VVK}. This numerical growth 
 of the coefficients is on top of $n!$ and $m!$ factors in the expressions for the amplitudes \eqref{Ah-fin}-\eqref{AZ-fin}. 

\subsection{Amplitudes above the threshold}

The operator $d_t^2 +M^2$ on the left hand side of the equations of motion away from the threshold
becomes $P_{\rm in}^2 +M^2$. Using the expression for the incoming momentum in \eqref{eqn:thrp}
together with the rescaled dimensionless variables\footnote{Note that the number of vector bosons is $m=2k$ for the amplitude 
${\cal A}_{h^* \to n\times h + 2k\times V_L}$, and $m=2k+1$ for the amplitude ${\cal A}_{V_L^* \to n\times h + (2k+1)\times V_L}$.}
\eqref{eq:kinHV} this amounts to the substitutions
\begin{eqnarray}
-\, (d_t^2 +1) \phi &\Longrightarrow& \left[\left(n(1+\varepsilon_h) +2k\, \kappa\,(1+\varepsilon_V)\right)^2-1\right]
\phi(n,2k)
\,,\label{2cleq-h3r}\\
-\, (d_t^2 +\kappa^2)  {\rm A} &\Longrightarrow &  \left[\left(n(1+\varepsilon_h) +(2k+1)\, \kappa\,(1+\varepsilon_V)\right)^2-\kappa^2\right]
{\rm A}(n,2k)\,,
\label{2cleq-A3r}
\end{eqnarray}
on the left hand sides of Eqs.~\eqref{cleq-h3r} and \eqref{cleq-A3r} respectively.
The quantities $\phi (n,2k)$ and ${\rm A}(n,2k)$ appearing on the right hand sides are the Taylor coefficients of the amplitudes
including the dependence on the external Higgs and the vector bosons kinematics. Based on the results of section 2,
we expect that in the large multiplicity limit $n,2k \to \infty$ with $n\varepsilon_h$ and $2k\varepsilon_V$ held fixed,
\begin{eqnarray}
\phi (n,2k) \,=\, d(n,2k) \exp\left[-C_h\,n\, \varepsilon_h \,-\, C_V\,2k\, \varepsilon_V\right]\,,
\label{eqCh} \\
{\rm A} (n,2k) \,=\, a(n,2k) \exp\left[-C_h\,n\, \varepsilon_h \,-\, C_V\,2k\, \varepsilon_V\right]\,.
\label{eqCV}
\end{eqnarray}
These expressions correspond to a rescaling 
 $z \to\, z\, e^{-C_h}$ and $W \to\, W\, e^{-C_V}$
in the Taylor expansions for the generating functions in \eqref{eq:dThA-finF}. As we already noted in
section 2,  in the large-multiplicity limit at hand, these expressions with the rescaled variables satisfy the full recursion relations for 
the amplitudes on and above the threshold. But in order to determine the values of the constants $C_h$ and $C_V$ 
we should first consider the $\varepsilon_h \to 0$ and $\varepsilon_V \to 0$ limit at finite general values of $n$ and $2k$.

We will first set $2k=0$ and expand \eqref{eqCh}-\eqref{eqCV} to the first order in $\varepsilon_h$ as follows,
\[
\phi (n,0) \,=\, d(n,0) + \varepsilon_h f(n,0)\,, \quad
{\rm A} (n,0) \,=\, a(n,0) + \varepsilon_h b(n,0)\,,
\label{eq:afk0}
\]
making no a priori assumptions about the form of $f(n,0)$ and $b(n,0)$, and solving for these coefficients
to determine the value of $C_h$ and the applicability of \eqref{eqCh}-\eqref{eqCV}.

Next, we will set $n=0$ and determine the constant $C_V$.

\subsubsection{n-Higgs production: Solving gauge and Higgs equations for 2k=0 for all n}

Here we consider the case where no additional gauge bosons were produced in the final state.
We thus set $2k=0$ and substitute for the amplitudes coefficients the leading-order expansion in terms 
of the Higgs kinetic energy, $d(n,0) + \varepsilon_h f(n,0)$ and 
$a(n,0) + \varepsilon_h b(n,0),$
\begin{eqnarray}
  {\cal A}_{h^* \to n\times h} (\varepsilon_h) &=& (2v)^{1-n}\, n!\, \left(d(n,0) +  \varepsilon_h f(n,0)\right)\,,
  \label{Ah-fink0}
\\
 {\cal A}_{Z_L^* \to n\times h + Z_L}(\varepsilon_h) &=& \frac{1}{(2v)^{n}}\, n!\, \left(a(n,0) +  \varepsilon_h b(n,0)\right)\,.
  \label{AZ-fink0}
\end{eqnarray}
Expanding the left hand sides of the classical equations \eqref{2cleq-h3r}-\eqref{2cleq-A3r} up to the order-$\varepsilon_h^1$ we have:
\begin{eqnarray}
 \left[n^2(1+\varepsilon_h)^2 -1\right]  
\left(d(n,0)+  \varepsilon_h f(n,0)\right)
&=& 
(n^2-1)\,d(n,0)
\label{2cleq-h3rk0}
\\
&+& \varepsilon_h^1\times\left[
(n^2-1)f(n,0) +2n^2d(n,0)
\right]
\nonumber
\,,
\\
 \left[\left(n(1+\varepsilon_h) +\kappa\right)^2-\kappa^2\right]
\left(a(n,0) +  \varepsilon_h b(n,0)\right)
&=& 
(n^2+2n\kappa)\,a(n,0)
\label{2cleq-A3rk0}
\\
&+& \varepsilon_h^1\times\left[
(n^2+2n\kappa)\,b(n,0) +2(n^2+n\kappa)\,a(n,0)
\right]
\nonumber
\end{eqnarray}
The $\varepsilon_h$-independent contributions give rise to the familiar equations for amplitudes on the threshold,
(these are Eqs.~\eqref{cleq-h3rT}-\eqref{cleq-A3rT} restricted to  $k=0$)
which have provided us with the expressions for $d(n,0)$ and $a(n,0)$.

The order-$\varepsilon_h^1$ contributions result in the equations for the off-threshold corrections to the amplitudes which
take the form,
\begin{eqnarray}
\frac{n-1}{n}
\left[
(n^2-1)f(n,0) +2n^2d(n,0)
\right]
 &=& \left.6\, \left( F \phi \,+\, F \phi^2\right)  \right|_{z^n W^0}\,,
\label{eq-hk0}\\
\frac{n-1}{n}
\left[
(n^2+2n\kappa)\,b(n,0) +2(n^2+n\kappa)\,a(n,0)
\right]
 &=& \left.
4\kappa^2 \left[(F+2F \phi)\, {\rm A} + (\phi+\phi^2)\,B
 \right]\right|_{z^n W^0}
 \,,
 \nonumber\\
 \label{eq-Ak0}
\end{eqnarray}
where  the function $F(z)$ is the same as in \eqref{eq:Fz},
and we have similarly defined  the new function $B(z)$ appearing in \eqref{eq-Ak0} via:
\[
F(z) \,=\, \sum_{n=2}^\infty\, \frac{n-1}{n} \, f(n,0) \, z^n\,, \quad
B(z) \,=\, \sum_{n=2}^\infty\, \frac{n-1}{n} \, b(n,0) \, z^n\,.
\label{eq:BFz}
\]

The equations \eqref{eq-hk0}-\eqref{eq-Ak0} are the recursion relations for the
coefficients $f(n,0)$ and $b(n,0)$ which determine the amplitudes' dependence on the kinematics.
These equations are obtained following the same prescription as we have used in deriving \eqref{eq:AmpsNepsR} in the scalar theory.
In fact, the first equation \eqref{eq-hk0} is identical to \eqref{eq:AmpsNepsR}
as it does not contain gauge-fields
in the $W^0$ selection rule we have imposed\footnote{
The  the last term on the right hand side of 
containing the gauge field ${\rm A}$ vanishes for $W=0$.}.

Specifically, the left hand sides of the equations \eqref{eq-hk0}-\eqref{eq-Ak0} are given by the
order-$\varepsilon_h^1$ contributions to the kinetic terms \eqref{2cleq-h3rk0}-\eqref{2cleq-A3rk0}
times the overall factor of $\frac{n-1}{n}$ appearing for the same reason as in \eqref{eq:AmpsNeps}.
The expressions on the right hand side in  \eqref{eq-hk0}-\eqref{eq-Ak0} arise as the 
leading order $\varepsilon_h^1$ expansion of the expressions on the right hand side of \eqref{cleq-h3rT}-\eqref{cleq-A3rT}
with the  substitution $\phi \to d(n,0) + \varepsilon_h f(n,0)$ and 
${\rm A} \to a(n,0) + \varepsilon_h b(n,0),$ and again accompanied by the relevant $\frac{n-1}{n}$ factors, as reflected
in the definitions \eqref{eq:BFz}.

The scalar field equation \eqref{eq-hk0} was solved in section 2 with the solution given by \eqref{eq:resultf}.
We can now solve the equation \eqref{eq-Ak0} for the gauge field recursively with {\it Mathematica}.
We thus determine the coefficients $b(n,0)$ for all values of $n\ge 2$.
In the large-$n$ limit we find that solving each of the equations result in the same leading-order behaviour 
for the coefficients,
\[
\frac{f(n,0)}{d(n,0)} \,\to\, -\, \frac{7}{6}\, n\,, \qquad
\frac{b(n,0)}{a(n,0)} \,\to\, -\, \frac{7}{6}\, n\,.
\label{eq:resultk0}
\]
This result (which we also checked does not depend on the value of $\kappa$) is not a coincidence.
Given the $n$-Higgs amplitude behaviour which we derived in \eqref{eq:expsc},
it must be the case that in the double-scaling $n \varepsilon_h=$fixed large-$n$ limit
the amplitudes \eqref{Ah-fink0}-\eqref{AZ-fink0}
exponentiate
\begin{eqnarray}
  {\cal A}_{h^* \to n\times h} (\varepsilon_h) &=& (2v)^{1-n}\, n!\, d(n,0)\exp\left[-\frac{7}{6}\,n\, \varepsilon_h \right]\,,
  \label{Ah-fink0F}
\\
 {\cal A}_{Z_L^* \to n\times h + Z_L}(\varepsilon_h) &=& \frac{1}{(2v)^{n}}\, n!\, a(n,0) \exp\left[-\frac{7}{6}\,n\, \varepsilon_h \right]
 \,.
  \label{AZ-fink0F}
\end{eqnarray}
These expressions for the amplitudes away from the threshold correspond to the rescaling $z \to\, z\, e^{-\frac{7}{6}}$
in the Taylor expansions for the generating functions in \eqref{eq:dThA-finF}. 

Our next goal is to determine the constant $C_V$ in the  exponential factor for the amplitudes  \eqref{eqCh}-\eqref{eqCV}
when the gauge bosons are present in the final state.

\subsubsection{2k-Vector production: Solving gauge and Higgs equations for n=0 for all k}

We now consider the case where only the vector bosons are produced in the final state, thus we keep $2k$
general and set $n=0$.
The equations \eqref{eqCh}-\eqref{eqCV} are expanded to the first order in $\varepsilon_V$ 
for $\varepsilon_h=0$. We have,
\[
\phi (0,2k) \,=\, d(0,2k) + \varepsilon_V\, f(0,2k)\,, \quad
{\rm A} (0,2k) \,=\, a(0,2k) + \varepsilon_V\, b(0,2k)\,,
\label{eq:afn-}
\]
once again, making no a priori assumptions about the form of $f(0,2k)$ and $b(0,2k)$, and solving for these coefficients
to determine the value of $C_V$ in \eqref{eqCh}-\eqref{eqCV}.

Repeating the same steps as in the previous sub-section we can write down the recursion relations
at the order-$\varepsilon_V^0$ ({\it cf.}  \eqref{cleq-h3rT}= \eqref{cleq-A3rT}):
\begin{eqnarray}
\left[(2k\, \kappa )^2 -1\right]\, d(0,2k)  &=& \left.\left[
3 \phi^2 + 2\phi^3 +2\kappa^2\left(1 +2\phi\right)W {\rm A}^2
\right]\right|_{z^0 W^k}\,,
\label{eq:thrdn0}\\
\left[(1+ 2k )^2 -1\right]\, a(0,2k)  &=& \left.
4 \left[ (\phi+\phi^2)\, {\rm A}
 \right]\right|_{z^0 W^k}
 \,,
\label{eq:thran0}
\end{eqnarray}
and at the order-$\varepsilon_V^1$ the Higgs-field equation is,
\begin{eqnarray}
\frac{2k-1}{2k}
\left[
((2k\kappa)^2-1)f(0,2k) + 8(k\kappa)^2 d(0,2k)
\right]
\, =
 \nonumber
 \\
\left[6 F \phi(1+ \phi) 
 + 4\kappa^2 F W {\rm A}^2
 + \left. 
4\kappa^2\left(1 +2\phi\right)W {\rm A} B
 \right]  \right|_{z^0 W^k}\,,
\label{eq-hn0}
\end{eqnarray}
and the gauge-field equation is,
\begin{eqnarray}
\frac{2k-1}{2k}
\left[
4k(k+1)\,b(0,2k) +2(k+1)^2\,a(0,2k)
\right]
= \left.
4 \left[ (F+2\phi F)\, {\rm A}
+(\phi+\phi^2)\, B
 \right]\right|_{z^0 W^k}
 \label{eq-An0}
\end{eqnarray}
The functions $F(W)$ and $B(W)$ are defined here  via:
\[
F(W) \,=\, \sum_{k=2}^\infty\, \frac{2k-1}{2k} \,\, f(0,2k) \, W^k\,, \quad
B(W) \,=\, \sum_{k=2}^\infty\, \frac{2k-1}{2k} \,\, b(0,2k) \,W^k\,.
\label{eq:BFW}
\]

We have solved numerically the order-$\varepsilon_V^0$ equations \eqref{eq:thrdn0}-\eqref{eq:thran0} 
by iterations with {\it Mathematica} to determine the coefficients $d(0,2k)$ and $a(0,2k)$.
Using these we solved the  order-$\varepsilon_V^1$ equations \eqref{eq-hn0}-\eqref{eq-An0} for the
coefficients $f(0,2k)$ and $b(0,2k)$ for different numerical values of the $\kappa$ parameter.

In the large-multiplicity limit $2k \to \infty$ our numerical results for the ratios of the coefficients,
\[
\frac{f(0,2k)}{d(0,2k)} \,\to\, -\, C_V(\kappa) \, 2k\,, \qquad
\frac{b(0,2k)}{a(0,2k)} \,\to\, -\, C_V(\kappa) \, 2k\,,
\label{eq:resultk01}
\]
confirm that both ratios: for the scalar-field coefficients, and for the gauge-field coefficients 
approach the same numerical constant $C_V$, which itself depends on the value of the mass-parameter $\kappa$.
For the physical value $\kappa= M_W/M_h= 80.384/125.66\simeq 0.64$, we get
\[ C_V \simeq 1.7 \,, \quad {\rm for} \quad \kappa = 0.64\,.
\label{eq:CVresult}
\]
More generally, defining
\begin{eqnarray}
-\,\frac{1}{2}\left(\frac{f(0,2k)}{d(0,2k)}-\frac{f(0,2k-2)}{d(0,2k-2)}\right) \,\to\,  C^{\,\rm scal.}_V(\kappa)\,, \quad
-\,\frac{1}{2}\left(\frac{b(0,2k)}{a(0,2k)}-\frac{b(0,2k-2)}{a(0,2k-2)}\right) \,\to\,  C^{\,\rm vect.}_V(\kappa)\,,
\nonumber
\end{eqnarray}
we get with $2k=54$,
\begin{eqnarray}
 C^{\,\rm scal.}_V \simeq 3.342
 \,, \qquad 
  C^{\,\rm vect.}_V \simeq 3.336
\,,  \quad &&{\rm for} \quad \kappa = 0.55 
 \nonumber\\
 C^{\,\rm scal.}_V \simeq 1.702
 \,, \qquad 
  C^{\,\rm vect.}_V \simeq 1.696
\,,  \quad &&{\rm for} \quad \kappa = 0.64 
\nonumber \\
 C^{\,\rm scal.}_V \simeq 0.996
 \,, \qquad 
  C^{\,\rm vect.}_V \simeq 0.996
\,,  \quad &&{\rm for} \quad \kappa = 1.
 \nonumber\\
 C^{\,\rm scal.}_V \simeq 0.829
 \,, \qquad 
  C^{\,\rm vect.}_V \simeq 0.829
\,,  \quad &&{\rm for} \quad \kappa = 2.
\nonumber \\
C^{\,\rm scal.}_V \simeq 0.805
 \,, \qquad 
  C^{\,\rm vect.}_V \simeq 0.805
\,,  \quad &&{\rm for} \quad \kappa = 3. 
 \nonumber\\
 C^{\,\rm scal.}_V \simeq 0.794
 \,, \qquad 
  C^{\,\rm vect.}_V \simeq 0.794
\,,  \quad &&{\rm for} \quad \kappa = 5.
\nonumber \\
 C^{\,\rm scal.}_V \simeq 0.789
 \,, \qquad 
  C^{\,\rm vect.}_V \simeq 0.789
\,,  \quad &&{\rm for} \quad \kappa = 10.
 \nonumber\\
 C^{\,\rm scal.}_V \simeq 0.787
 \,, \qquad 
  C^{\,\rm vect.}_V \simeq 0.787
\,,  \quad &&{\rm for} \quad \kappa = 100.
\end{eqnarray}
The convergence of the series is improved at higher values of $\kappa$.
A very likely conclusion is that at an unphysical value of $\kappa=1$ which corresponds to $M_h=M_V$ the
value of the vector constant is $C_V=1.$
\medskip

\noindent Our final result for the tree-level multi-vector-boson multi-Higgs production amplitudes in the
high-multiplicity double-scaling limit for $\kappa= M_W/M_h= 80.384/125.66\simeq 0.64$ is: 
\begin{eqnarray}
{\cal A}_{h^* \to n\times h + m\times Z_L} &=& (2v)^{1-n-m}\, n!\, m! \, d(n,m)
\, \exp\left[-\,\frac{7}{6}\,n\, \varepsilon_h - 1.7 \,m\, \varepsilon_V\right]\,,
  \label{Ah-finkn}
\\
{\cal A}_{Z_L^* \to n\times h + (m+1)\times Z_L} &=& \frac{1}{(2v)^{n+m}}\, n!\, (m+1)!\, a(n,m)
 \,  \exp\left[-\,\frac{7}{6}\,n\, \varepsilon_h - 1.7 \,m\, \varepsilon_V \right]
  \label{AZ-finkn}
\end{eqnarray}

The expressions \eqref{Ah-finkn}-\eqref{AZ-finkn} constitute our main results, as far as the scattering amplitudes 
at high multiplicities are concerned.
They incorporate the dependence on the momenta of final particles (computed in the 
$n\, \varepsilon_h \,+\,\,m\, \varepsilon_V $ = fixed regime) and hence can be integrated over the phase-space.
The lesson we draw from the amplitude expressions above is that even away from the multi particle thresholds they mainain 
the factorial dependence on the multiplicities and the further enhancement by the growing coefficients $d(n,m)$ and
$a(n,m)$, found in the threshold amplitudes. The dependence on the kinematic variables of the final space provides only a mild
suppression of the result on the threshold -- at least in the regime where the derivation of \eqref{Ah-finkn}-\eqref{AZ-finkn}
is valid. For example, for $n=30$ and $\varepsilon_h=0.1$ so that $n\varepsilon_h=3$ is an order-one constant,
we have $ e^{-\,\frac{7}{6}\,n\, \varepsilon_h } \simeq 0.03$ and similarly
for $m=30$ and $\varepsilon_V=0.1$ we have $ e^{- 1.7 \,m\, \varepsilon_V  } \simeq 0.006$ as overall multiplicative
factors in the amplitudes. 

In the next section we will integrate these amplitudes over the phase space in order
to estimate the rates for these processes.

\medskip
\section{Integrating over the phase space}
\label{sec:four}
\medskip

The scattering cross sections for multi-particle production rates arise from integrating the squared amplitudes 
\eqref{Ah-finkn}-\eqref{AZ-finkn} over the Lorentz-invariant phase space,
\[
\sigma_{n,m} \,=\, \int d\Phi_{n,m}\, \frac{1}{n!\,m!} \, \left|{\cal A}_{h^* \to n\times h + m\times Z_L}\right|^2\,,
\label{eq:sigmanm}
\]
where $1/n!$ and $1/m!$ are the Bose statistics factors accounting for the $n$ identical Higgses and $m$ identical
longitudinal vector boson states, and we have dropped the overall flux factor on the {\it r.h.s.} of \eqref{eq:sigmanm}.
The next step is to integrate over phase space. The $n$-particle Lorentz-invariant phase space volume element 
has the familiar form,
\[ 
\int d\Phi_{n} \,=\, (2\pi)^4 \delta^{(4)}(P_{\rm in}-\sum_{j=1}^n p_j) \,
 \prod_{j=1}^n \int \frac{d^3 p_j}{(2\pi)^3\, 2 p_j^0} \,,
\]
but in order to use in \eqref{eq:sigmanm} our results for the amplitudes \eqref{Ah-finkn} within their
region of validity -- i.e. the high-multiplicity non-relativistic limit -- the phase space integrations have to be performed 
in the same non-relativistic approximation.

We note that it should not come as a surprise that the large-$n$ small-$\varepsilon$ limit will amount to a very small 
phase-space volume. Indeed, a very rough estimate for the phase-space volume in this approximation 
will be $\Phi_n \propto M^{3n} \times \varepsilon^{3n/2}$. In dimensionless units, it arises from the 
product of $n$ three-dimensional spherical volumes obtained by integrating over each of the final particle momenta 
$|p|_i \lesssim M\sqrt{2\varepsilon}$. It is then not surprising that the resulting volume of the non-relativistic
$n$-particle phase-space reduces the cross section by the factor $\propto \varepsilon^{3n/2}$ which is $\ll 1$ in the
limit $\varepsilon \to 0$ and $n\gg 1$. We will confirm this estimate with a more precise computation below, but it is important to stress
from the outset that the suppression of the resulting cross sections at moderate energies is entirely caused by the
non-relativistic approximation used in computing the phase-space volume, and is not driven by the form of the
amplitudes squared. In order to compute the rate in the more realistic settings, one should integrate over a larger portion of the phase-space.
In the present paper we will not pursue this route as this would require knowing the amplitudes beyond the non-relativistic limit.

The phase-space integration in the large-$n$ non-relativistic limit with $n \varepsilon_h$ fixed is easily carried out 
by integrating over
 the $d^{3n}p$  volume of the $3n$-dimensional  of radius $|p| = M_h \sqrt{2n \varepsilon_h}$. The resulting
 non-relativistic phase space volume in the large-$n$ limit is (see e.g. \cite{Son}),
\[ 
\Phi_n \,\simeq\, 
   \frac{1}{\sqrt{n}} \left(\frac{M_h^2}{2}\right)^n \, \exp\left[ \frac{3n}{2}\left(\log \frac{\varepsilon_h}{3\pi} +1 \right)
  \,+\, \frac{n \varepsilon_h}{4} \,+\, {\cal O}(n \varepsilon_h^2)
  \right]\,.
\]
Combining this with the $n$-Higgs amplitude squared (with $m=0$ vector bosons), we get,
\begin{eqnarray}
&&\frac{1}{n!} \, \Phi_n  \, \left|{\cal A}_{n}\right|^2 \,\simeq \, \Phi_n \,
(2v)^{-2n}\, n! \,  d(n,0)^2
\, \exp\left[-\,\frac{7}{3}\,n\, \varepsilon_h \right]    
\nonumber \\
&&\quad \simeq   \frac{1}{\sqrt{n}}\,
 \exp\left[ 2 \log d(n,0)\,+\,n\left(\log \frac{\lambda n}{4} -1 \right)
 \,+\,\frac{3n}{2}\left(\log \frac{\varepsilon_h}{3\pi} +1 \right)
 \, -\,\frac{25}{12}\,n\, \varepsilon_h \right]  
\end{eqnarray}

Repeating the same steps for vector boson emissions we now can write down the rate for the high multiplicity
$n$-Higgs + $m$-vector boson production corresponding to the  amplitude \eqref{Ah-finkn},
\begin{eqnarray}
&&\sigma_{n,m} \,\sim\, \exp \left[ 2 \log d(n,m)\,+\,n\left(\log \frac{\lambda n}{4} -1 \right) \,+\, 
m \left(\log\left(\frac{g^2 m}{32}\right) - 1\right) \right.
\label{eq:FnmH} \\
&&\quad+ \left. \frac{3n}{2}\left(\log \frac{\varepsilon_h}{3\pi} +1 \right)
 \,+\,\frac{3m}{2}\left(\log \frac{\varepsilon_V}{3\pi} +1 \right)
 \, -\,\frac{25}{12}\,n\varepsilon_h  \, -\,3.15 \,m\varepsilon_V\,+\, {\cal O}(n\varepsilon_h^2+m\varepsilon_V^2)
 \right]  \nonumber
\end{eqnarray}
The cross section arising from the amplitude \eqref{AZ-finkn} takes the same form as \eqref{eq:FnmH} but with the
$2 \log a(n,m)$ factor on the right hand side. The numerical coefficients $d(n,m)$ and $a(n,m)$ were derived in 
\cite{VVK} by solving recursion relations for the amplitudes on the multi-particle mass threshold; they are plotted in Fig.~\ref{fig:da}.
 \begin{figure}[t]
\begin{center}
\begin{tabular}{cc}
\hspace{-.4cm}
\includegraphics[width=0.5\textwidth]{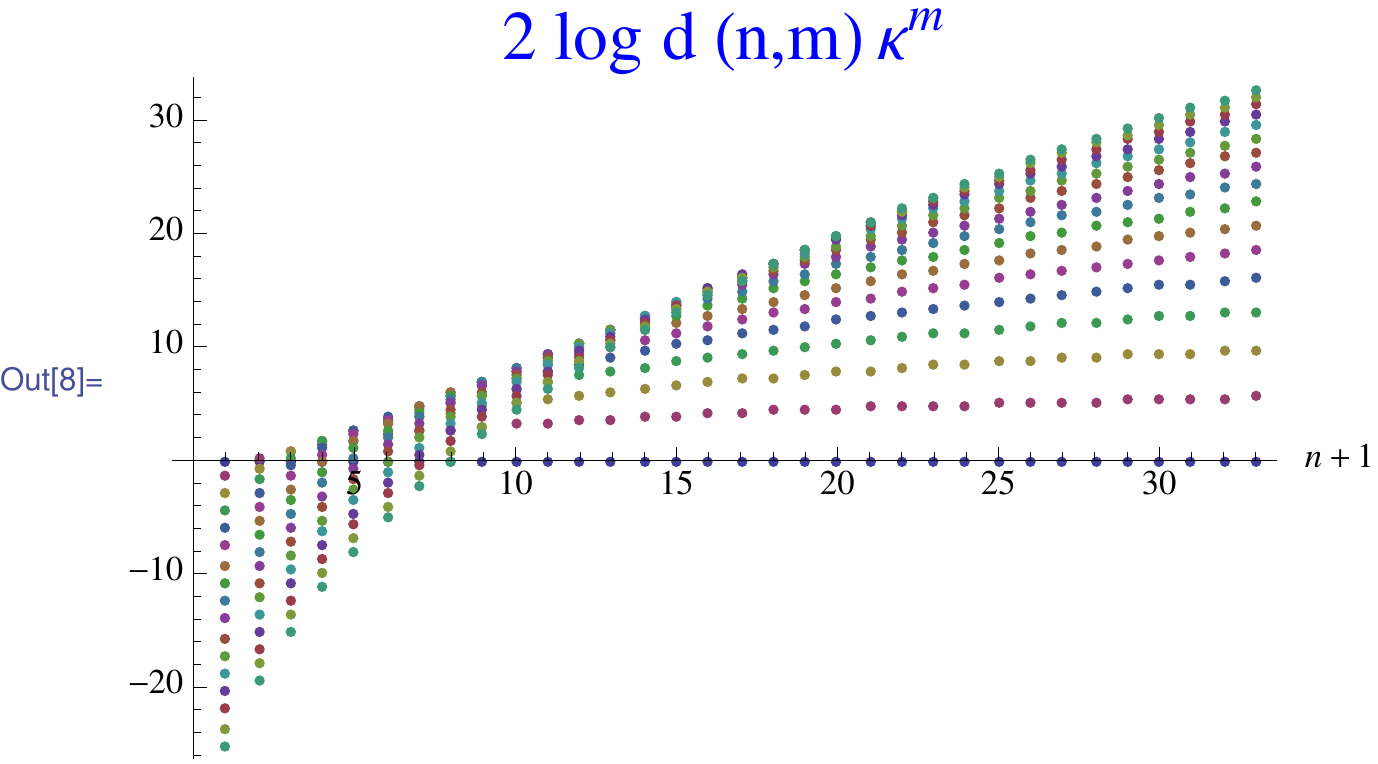}
&
\includegraphics[width=0.5\textwidth]{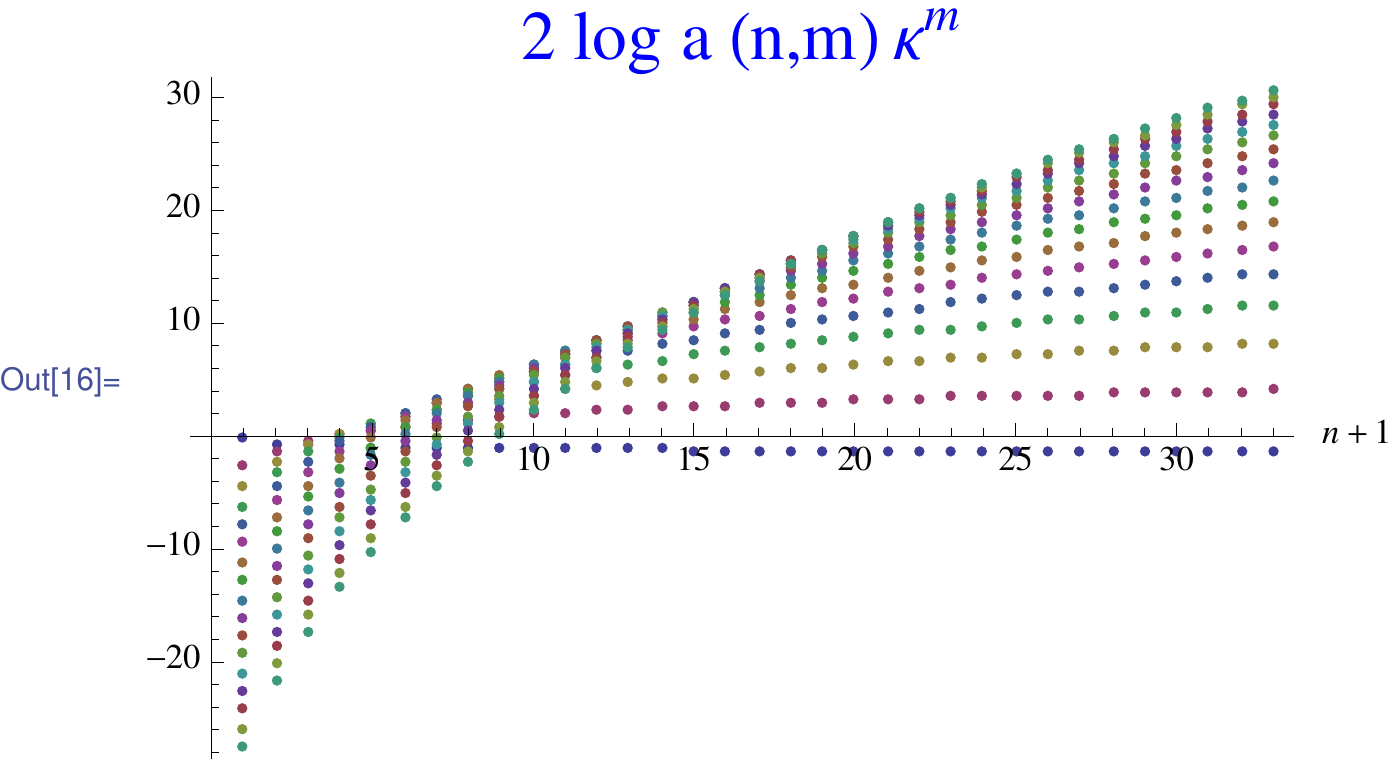}
\\
\end{tabular}
\end{center}
\vskip-.4cm
\caption{
Amplitude coefficients of Figure~\ref{fig:da}  in the form $2 \log (\kappa^m d(n,m))$ and $2 \log (\kappa^m a(n,m))$
appearing in Eqs.~\eqref{eq:FnmH2}-\eqref{eq:FnmH3}. The label $n=0,1,\ldots, 32$ is shown along the horizontal axis and the sequence of curves corresponds to 
$m=0, 2, \ldots,32$ with $m$ increasing from bottom to top (on the right of each plot). 
}
\label{fig:daK}
\end{figure}

At $m=0$ all $d$-coefficients are equal to one, hence the first term on the right hand side vanishes in this case, $2 \log d(n,m=0) = 0$.
At higher values of $m$, however the coefficients $d(n,m)$ and  $a(n,m)$ start growing. 
To somewhat tame the numerical growth of the Taylor coefficients we can rescale them with a factor of $\kappa^m$ and this
can be nicely combined with the observation that $m\log(\frac{g^2 m}{32}) =m\log (\kappa^2) +m\log( \frac{\lambda m}{4})$
which facilitates 
a re-write of \eqref{eq:FnmH} in the form:
\begin{eqnarray}
&&\sigma_{n,m} \,\sim\, \exp \left[ 2 \log (\kappa^m d(n,m))\,+\,n \log \frac{\lambda n}{4}\,+\, 
m \log \frac{\lambda m}{4}\right.
\label{eq:FnmH2} \\
&&\quad+ \left. \frac{n}{2}\left(3\log \frac{\varepsilon_h}{3\pi} +1 \right)
 \,+\,\frac{m}{2}\left(3\log \frac{\varepsilon_V}{3\pi} +1 \right)
 \, -\,\frac{25}{12}\,n\,\varepsilon_h  \, -\,3.15\,m\,\varepsilon_V\,+\, {\cal O}(n\varepsilon_h^2+m\varepsilon_V^2)
 \right]  \nonumber
\end{eqnarray}
The amplitudesTaylor coefficients in the form $2 \log (\kappa^m d(n,m))$ and $2 \log (\kappa^m a(n,m))$
appearing on the right hand side are shown in Fig.~\ref{fig:daK}.

\medskip
\section{Conclusions}
\label{sec:concl}
\medskip

The expressions in Eq.~\eqref{eq:FnmH} or equivalently in Eq.~\eqref{eq:FnmH2}
characterise the cross section $\sigma_{n,m}$ for the multi-particle $n$-Higgs $m$-vector boson 
production obtained in the high-multiplicity non-relativistic limit.
They were derived in the Gauge--Higgs theory and based on computing all 
tree-level scattering amplitudes with $n$-Higgs $m$-longitudinal-vector boson 
final states, derived on- and off- the multi-particle mass threshold in Eqs.~\eqref{Ah-finkn}-\eqref{AZ-finkn}.
These are our main results.

\bigskip

As we have already noted, the imposition of the non-relativistic limit dramatically reduces the 
otherwise available phase-space
to a tiny volume $\propto \exp\left[ -\frac{3n}{2} \log \frac{3\pi/e}{\varepsilon_h} - \frac{3m}{2} \log \frac{3\pi/e}{\varepsilon_V }\right].$
At moderately high energies this suppression factor will dominate the cross section, as we will illustrate below.
However, one should  keep in mind that this effect is is simply an artifact of the approximation used for computing the phase-space.

\bigskip

Before we conclude, it will be useful to sketch some simple estimates for the energy scales involved.
First, we would like to estimate the value of  $\log (\sigma_{n,m})$  for a ``minimal interesting value" 
of final particle multiplicities, $n=m=30$ which is roughly $1/\alpha_W$ (they of course also correspond to the highest
multiplicities where we have calculated the values of the Taylor coefficients $d(n,m)$ and $d(n,m)$).
Following our expression in Eq.~\eqref{eq:FnmH2} we have,
\begin{eqnarray}
n=m \,=\, 30  &\Longrightarrow& 2 \log (\kappa^m d(n,m))\,\simeq\, 30\,,
\\
\sqrt{\hat{s}}\, \simeq \, 6.8\, {\rm TeV} && n \log \frac{\lambda n}{4} \,=\, m \log \frac{\lambda m}{4} \,\simeq \, -0.02\,\simeq\, 0\,,
\label{eq:logdresc}
\\
\nonumber
\\
\varepsilon_h=\varepsilon_V \,=\, 0.1  &\Longrightarrow& 
 \frac{n}{2}\left(3\log \frac{\varepsilon_h}{3\pi} +1 \right) \,=\, \frac{m}{2}\left(3\log \frac{\varepsilon_V}{3\pi} +1 \right)\,\simeq\, -\,190
 \,, \label{eq-nrelsel} 
 \\
&&  -\,\frac{25}{12}\,n\varepsilon_h  \, -\,3.15\,m\varepsilon_V \,\simeq\, -\,15\,,
\end{eqnarray}
where we have been careful in \eqref{eq-nrelsel} to select an appropriately small value 0.1 of the kinetic energy per particle to be
consistent with the non-relativistic limit.
This amounts to
\[
n=m \simeq 30  \,\Longrightarrow\,\qquad
 \log (\sigma) \,\simeq\, 30\,-\, 190\,-\, 190\,-\,15\,=\, -365\,,
\]
which amounts to a negligibly small rate $\sigma_{30,30}  \,\simeq\, 0.3 \times 10^{-160} .$ Clearly, to have a higher rate,
we need to increase the number of particles in the final state. 

\medskip

Still, at even higher multiplicities perturbation theory will break down and perturbative unitarity
will be violated by exponentially growing rates even within the current non-relativistic phase-space limit. To see this, let us re-arrange the expression in Eq.~\eqref{eq:FnmH2} as follows:
\begin{eqnarray}
&&\sigma_{n,m} \,\sim\, \exp \left[ 2 \log (\kappa^m d(n,m))\,+\,
m\,\log \left(\frac{\lambda\, m\varepsilon_V}{12\pi}\sqrt{\frac{\varepsilon_V}{3\pi}}\right)  \,+\,m\left(0.5 -\,3.15\,\varepsilon_V\right) \right.
\nonumber\\
&& \qquad\qquad\qquad+ \left. 
n\,\log \left(\frac{\lambda\, n\varepsilon_h}{12\pi}\sqrt{\frac{\varepsilon_h}{3\pi}}\right) 
\,+\, 
n\left(\frac{1}{2} -\,\frac{25}{12}\,\varepsilon_h\right)
 \,+\, {\cal O}(n\varepsilon_h^2+m\varepsilon_V^2)
 \right]  \,.
 \label{eq:FnmH3} 
\end{eqnarray}
This result holds in the double scaling limit, $n\to \infty$, $m\to \infty$, $\varepsilon_h\to 0$, $\varepsilon_V\to0$
with $n\varepsilon_h$ and $m\varepsilon_V$ held fixed.

\medskip

We now consider a somewhat extreme case with the number of produced vector bosons is very large, $m \simeq 7500$
and we keep the number of Higgs bosons small, for simplicity. Then only the terms on the first line of Eq.~\eqref{eq:FnmH3}  matter.
If we also assume $\varepsilon_V =0.5$ we would get,
\begin{eqnarray}
m \,=\, 7500 \,, \,\,  \varepsilon_V \,=\, 0.5 &\Longrightarrow& 
m\,\log \left(\frac{\lambda\, m\varepsilon_V}{12\pi}\sqrt{\frac{\varepsilon_V}{3\pi}}\right)  \,+\,m\left(0.5 -\,3.15\,\varepsilon_V\right)
\,\simeq\, 0\,,
\\
\nonumber \\
 \sqrt{\hat{s}}\, \simeq \, 845\, {\rm TeV}\, &&
 \sigma_{n,m} \,\sim\, \exp \left[ 2 \log (\kappa^m d(n,m)) \right]  \,,
\end{eqnarray}
and if we increase the number of vector bosons by 100 more at a cost of extra 10 TeV at these energies, 
we would get 
\begin{eqnarray}
m \,=\, 7600 \,, \,\,  \varepsilon_V \,=\, 0.5 &\Longrightarrow& 
m\,\log \left(\frac{\lambda\, m\varepsilon_V}{12\pi}\sqrt{\frac{\varepsilon_V}{3\pi}}\right)  \,+\,m\left(0.5 -\,3.15\,\varepsilon_V\right)
\,\simeq\, 102\,,
\\
\nonumber \\
 \sqrt{\hat{s}}\, \simeq \, 855\, {\rm TeV}\, &&
 \sigma_{n,m} \,\sim\, \exp \left[ 2 \log (\kappa^m d(n,m)) \right] \times e^{ 102} \ggg 1\,,
\end{eqnarray}
This behaviour is obviously in violation of perturbative unitarity even if we do not worry about the additional factor of $2 \log (\kappa^m d(n,m)) $
which is likely to continue growing beyond the value of 30 in \eqref{eq:logdresc} at these multiplicities. 
This regimes is also beyond the validity region of Eq.~\eqref{eq:FnmH3}  since the unknown corrections of the order $m \varepsilon_V^2$
are large.

Similarly, in the  case of mostly Higgs production, i.e. at low $m$ and $n\simeq 4000$ we find,
\begin{eqnarray}
n \,=\, 4000 \,, \,\,  \varepsilon_h \,=\, 0.65 &\Longrightarrow& 
n\,\log \left(\frac{\lambda\, n\varepsilon_h}{12\pi}\sqrt{\frac{\varepsilon_h}{3\pi}}\right)  \,+\,n\left(1/2 -\,25/12\,\varepsilon_h\right)
\,\simeq\, 23\,,
\\
\nonumber \\
 \sqrt{\hat{s}}\, \simeq \, 830\, {\rm TeV}\, &&
 \sigma_{n,m} \,\sim\, \exp \left[ 2 \log (\kappa^m d(n,m)) \right] \times e^{ 23} \ggg 1\,,
\end{eqnarray}

\medskip

The main conclusion we want to draw from the computations presented in this paper is that perturbation theory does break down 
in the weak sector of the Standard Model. This breakdown occurs already at leading order (i.e. tree-level) in the perturbative expansion.\footnote{The higher-loop corrections are expected to make this worse by introducing corrections of the order $\lambda n^2/(4\pi)$ and 
$ \alpha_W m^2/(4\pi)$ which are $\gg 1$ at the sufficiently high multiplicities.}
To accurately determine {\it the lower bound} on the energy scale where this breakdown does occur would require going beyond the 
double-scaling high-multiplicity non-relativistic limit we have assumed and used throughout (and would also require including higher-order corrections as well as
computing even higher multiplicity amplitudes). Our rough estimate is that the perturbative meltdown energy range is not far from 
a few hundred TeV even after including the effect of the highly suppressed non-relativistic phase-space volume.

To establish whether these very high multiplicity processes become observable and even 
dominant at future circular hadron colliders,
to determine what is the precise energy scale where this happens and what is the average number of bosons produced,
ultimately requires developing of a non-perturbative (possibly semi-classical) technique in the electro-weak sector of the Standard Model.
We plan to return to these open problems in future work.

\bigskip

\section*{Acknowledgements}

I am grateful to Joerg Jaeckel for illuminating discussions.
This work is supported by the UK Science and Technology Facilities Council through the IPPP grant, 
and by the  Royal Society Wolfson Research Merit Award.

\bigskip

%
\startappendix
\Appendix{Unbroken $\mathbold \phi^4$ theory}
\label{sec:app}
For completeness and to provide another illustration for using the formalism outlined in section 2, we
will re-derive here the results of Ref.~\cite{LRST} for multi-particle amplitudes in the 
$\phi^4$ theory with no spontaneous symmetry breaking,
\[
{\cal L} \,= \, \frac{1}{2} \left(\partial \phi\right)^2 - \frac{1}{2} M^2 \phi^2 - \frac{1}{4} \lambda \phi^4
 \,.
\label{Aeq:Lphi}
\]
The corresponding classical equation for the theory \eqref{Aeq:Lphi} is 
\[
-\, (\partial^\mu \partial_\mu +M^2)\, \phi \,=\,\lambda\,\phi^3\,,
\label{Aeq:varphi}
\]
and we are after the $n$-point amplitude in the non-relativistic limit $\varepsilon \ll 1$  in the form \eqref{eq:2-16},
\[
{\cal A}_{n} (p_1 \ldots p_n)  \,=\, n!\, (d_n \,+\, f_n \, \varepsilon)
\,.
\]
At the order $\varepsilon^0$ the classical equation \eqref{Aeq:varphi} gives the recursion relation for 
 $d_n$ ({\it cf.} Eq.~\eqref{eq:Amps3N}):
\[
(n^2 -1)\, d_{n}  \,=\, \frac{\lambda}{M^2} \sum_{n_1,n_2, n_3}^n  \delta^n_{n_1 +n_2+n_3} 
 d_{n_1} \, d_{n_2}  \, d_{n_3} \,, \qquad d_1=1\,,
\label{Aeq:Amps3N} \]
with the solution, $d_n=(\lambda/(8M^2))^{(n-1)/2}$ for $n=3,5,7,\ldots$.
At the order-$\varepsilon^1$ we can write down the recursion relation for the $f_n$ coefficients
following the same routine as we did in writing \eqref{eq:AmpsNeps},
\[
\frac{n-1}{n}\left(
\left( n^2-1 \right)  f_n \,+\, 2 n^2\,d_{ n} \right)
=  \, 3 \,\frac{\lambda} {M^2} \sum_{n_1,n_2, n_3}^n \delta^n_{n_1 +n_2+n_3} 
\frac{n_1-1}{n_1}\,
 f_{n_1}  \, d_{n_2} \, d_{n_3} \,.
 \label{Aeq:AmpsNeps}
\]
The factor of 3  on the right hand side of \eqref{Aeq:AmpsNeps}
accounts for the combinatorial factor which occurs in the order-$\varepsilon$ expansion 
of $(d_{n_1}+\varepsilon f_{n_1}) (d_{n_2}+\varepsilon f_{n_2}) (d_{n_3}+\varepsilon f_{n_3})$
to obtain $3\,\varepsilon\, f_{n_1}  \, d_{n_2} \, d_{n_3}$.
The factors of $\frac{n-1}{n}$ and $\frac{n_1-1}{n_1}$ on both sides of \eqref{Aeq:AmpsNeps} arise as before from
the form of kinetic energy in \eqref{KinEsubs}.

The recursion relation \eqref{Aeq:AmpsNeps} is solved by introducing the function $F(z)$, exactly as in 
\eqref{eq:AmpsNepsR}, and solving the ordinary differential equation
\[
\frac{n-1}{n}\left(
\left( n^2-1 \right)  f_n \,+\, 2 n^2\,d_{ n} \right)
\,=\,  3 \,\frac{\lambda} {M^2}\, F(z) \phi(z)^2|_{z^n}\,.
 \label{Aeq:AmpsNepsR}
\]
It is straightforward to solve this equation with {\it Mathematica} and we have checked that the solution is the same as found in 
\cite{LRST}, which reads
$f_n\,=\, - \, \left(\frac{5}{6} \,n \,-\, \frac{1}{6} \, \frac{n}{n-1}\right)\, d_n.$ 
The resulting amplitude in the double-scaling limit $n\to \infty$, $\varepsilon \to 0$, 
$n\varepsilon$ fixed,  is then given by
\[
{\cal A}_n^{\rm noSSB} \,=\, n!\,  \left(\frac{\lambda}{8M^2}\right)^{\frac{n-1}{2}}\exp\left[-\frac{5}{6}\,n\, \varepsilon \right]\,,
\]
in agreement with Ref.~\cite{LRST}.

\bigskip

\bibliographystyle{h-physrev5}

\end{document}